\documentclass{article}
\usepackage{graphicx}
\usepackage{xcolor}
\usepackage{geometry}
\usepackage{url}
\usepackage{pdfpages}
\usepackage{amsfonts}
\usepackage{booktabs}
\usepackage{multirow}
\usepackage{siunitx}
\usepackage{array}
\usepackage[normalem]{ulem}
\newcolumntype{L}[1]{>{\raggedright\let\newline\\\arraybackslash\hspace{0pt}}m{#1}}
\newcolumntype{C}[1]{>{\centering\let\newline\\\arraybackslash\hspace{0pt}}m{#1}}
\newcolumntype{R}[1]{>{\raggedleft\let\newline\\\arraybackslash\hspace{0pt}}m{#1}}

\usepackage{amsmath}

\title{A Multi-tasking Model of Speaker-Keyword Classification for Keeping Human in the Loop of Drone-assisted Inspection}

\author{%
    \textbf{Yu Li, Ph.D. Student}\\
  Department of Civil Engineering\\
  Stony Brook University, Stony Brook, NY 11794, USA\\
  Email: yu.li.5@stonybrook.edu\\
  ORCID=0000-0002-7245-0284\\
  \textbf{ Anisha Parsan, Research volunteer }\\
  Department of Civil Engineering\\
  Stony Brook University, Stony Brook, NY 11794, USA\\
\textbf{ Bill Wang,  Research volunteer }\\
  Department of Civil Engineering\\
  Stony Brook University, Stony Brook, NY 11794, USA\\
  \textbf{ Penghao Dong,   Ph.D. Student}\\
  Department of Mechanical Engineering\\
  Stony Brook University, Stony Brook, NY 11794, USA\\
    \textbf{Shanshan Yao,  Associate Professor}\\
  Department of Mechanical Engineering\\
  Stony Brook University, Stony Brook, NY 11794, USA\\
  ORCID=0000-0002-2076-162X\\
  \textbf{Ruwen Qin, Associate Professor}\\
  Department of Civil Engineering\\
  Stony Brook University, Stony Brook, NY 11794, USA\\
  Email: ruwen.qin@stonybrook.edu\\
  ORCID: 0000-0003-2656-8705
}
\date{}

\begin{document}
\maketitle 
\section{Abstract} 
Audio commands are a preferred communication medium to keep inspectors in the loop of civil infrastructure inspection performed by a semi-autonomous drone. To understand job-specific commands from a group of heterogeneous and dynamic inspectors, a model must be developed cost-effectively for the group and easily adapted when the group changes. This paper is motivated to build a multi-tasking deep learning model that possesses a Share-Split-Collaborate architecture. This architecture allows the two classification tasks to share the feature extractor and then split subject-specific and keyword-specific features intertwined in the extracted features through feature projection and collaborative training. A base model for a group of five authorized subjects is trained and tested on the inspection keyword dataset collected by this study. The model achieved a 95.3\% or higher mean accuracy in classifying the keywords of any authorized inspectors. Its mean accuracy in speaker classification is 99.2\%. Due to the richer keyword representations that the model learns from the pooled training data, adapting the base model to a new inspector requires only a little training data from that inspector, like five utterances per keyword. Using the speaker classification scores for inspector verification can achieve a success rate of at least 93.9\% in verifying authorized inspectors and 76.1\% in detecting unauthorized ones. Further, the paper demonstrates the applicability of the proposed model to larger-size groups on a public dataset. This paper provides a solution to addressing challenges facing AI-assisted human-robot interaction, including worker heterogeneity, worker dynamics, and job heterogeneity.

\hfill\break%
\noindent\textit{Keywords}: human-in-the-loop, human robot interaction, infrastructure inspection, keyword classification, speaker recognition 

\section{Introduction}

Safe, reliable civil infrastructure is a foundation for the nation’s socio-economic vitality. For example, the National Bridge Inventory has 619,588 bridges \cite{2022BridgeReport} spatially distributed on over 4,000,000 miles of public roads \cite{HighwayStat2020}. The average daily traffic passing the bridges is 4.627 billion \cite{StatusReport24}. However, 42\% of the bridges are over 50 years old, and over 55.1\% are rated as fair or poor \cite{2021ReportCard}, meaning they have deteriorated. Stakeholders closely monitor the health condition of bridges to assure the safety of passing traffic. In response to the vast demand for bridge inspection and because of the complexity of this mission, aerial robots such as drones have been introduced to improve the time efficiency, worker safety, and cost-effectiveness of inspection.

A human-robot system for a bridge inspection consists of an inspector and a drone. Their collaboration method directly impacts the system's job efficiency and task performance. For example, an inspector must possess the appropriate psychomotor, cognitive, and sensory abilities to operate the drone manually using a hand controller throughout the inspection  \cite{li2022virtual}. The drone is preferred to be at least semi-autonomous with the inspector's assistance or guidance in the loop. Specifically, the drone can automatically perform inspection tasks under predefined conditions, and the inspector will guide the drone or take control of it only when a need is identified. For example, the drone detects an area of concern nearby but off the pre-planned inspection path for a task. The drone hovers there and sends a message to the inspector. The inspector will judge and then tell the drone to continue its current task or guide it to add an incremental task. Human-robot interaction is essential when the robot is semi-autonomous with an inspector in the loop.

Some types of guidance that inspectors give to a semi-autonomous drone, such as triggering, terminating, and slightly modifying a task that the drone is performing automatically, can be provided conveniently using a set of commands. There are different media for communicating with the drone, such as speech commands, non-speech commands, remote controllers, and hand gestures. Speech commands have advantages over others because humans use them naturally in daily communication. Therefore, the mapping between speech commands and the drone's actions is intuitive to inspectors. A model is required to analyze the inspector's acoustic signals and classify the command keywords so that the drone can understand the inspector's guidance. Compared to the literature, the application to the collaborative human-robot inspection of bridges has unique characteristics or specifications. First, although only a small set of keywords is required, they are job-specific and not necessarily covered by existing big datasets of speech commands. Therefore, training and refining the model must be efficient, for instance, using a small sample of data collected for any inspection job. Second, a stakeholder such as a State department of transportation or a local transportation agency usually has a group of inspectors who differ in their background, acoustic characteristics, and speaking habits. The model must reliably recognize the speech commands for a group of heterogeneous inspectors, ranging from a few to tens. Third, the drone should only follow the instruction of authorized inspectors, not other workers at the inspection site or cyber attackers. That is, the model should be able to recognize and verify the inspector. Last, the model can adapt to workforce dynamics due to promotion, retirement, recruitment, and turnover. Robotic technology is proliferating in the industry, such as construction progress inspection, underwater infrastructure inspection, post-disaster search and rescue, aerial survey, military reconnaissance and surveillance, and others. The need for keeping humans in the loop of a semi-autonomous robot-assisted work and similar requirements for the speaker-keyword classification model signify the urgency of exploring a tailored solution. While the existing literature on speech command and speaker recognition address one or another need from some perspectives and in-depth, no focused study has specifically addressed those needs of this new engineering application.

Motivated by the above-discussed new capability that the collaborative human-robot inspection desires, this paper aims to develop a model that can reliably classify spoken keywords and determine who the speaker is if verified as an authorized inspector. Creating such a model for any group of inspectors and any inspection job must be cost-effective. When new inspectors join the group, refining the model to add additional classes of speakers is convenient. If any inspectors leave, their data still support the model and will not be obsolete. The remainder of the paper will detail the discussion. The next section summarizes the related work. Then, Section \ref{sec:Model} presents the proposed multi-tasking model, followed by the implementation details in Section \ref{sec:ImplementationDetails}. After that, Section \ref{sec:Results and Analysis} discusses results from experimental studies that demonstrate the model performance and determine requirements for achieving the performance. In the end, Section \ref{sec:Conclusions} concludes the study by summarizing research findings and important future work.

\section{The Literature}
\label{Sec:Literature}

The literature related to this paper includes keyword spotting or speech command recognition, acoustic signal-based speaker recognition, and multi-tasking models that integrate the two tasks into a unified model. 

\subsection{Keyword Spotting and Speech Command Recognition}

Speech command is one of the media to deliver human instruction to robots \cite{goodrich2008human}. Compared to other media such as hardware, gestures, and natural language, speech command is easier to implement. Developing a reliable recognition system for simple commands has had a sound methodological foundation. Therefore, various applications have chosen speech command as the communication media, such as smart homes \cite{arriany2016applying} and air traffic control \cite{holone2015possibilities}. Speech command recognition for keeping inspectors in the loop of a semi-autonomous drone-assisted civil infrastructure inspection has not been widely developed yet.

Along with the growing need for human-machine interaction, the development of lightweight models for recognizing simple commands is gaining growing interest. Keyword spotting is a small-scale speech recognition task identifying keywords from audio streams. Deep neural networks have recently outperformed standard Hidden Markov Models (HMMs) and become a new stream of speech command recognition methods \cite{chen2014small}. For example, convolutional neural networks (CNNs) designed for keyword spotting showed more than 95\% accuracy on the Google speech command dataset \cite{tang2018deep,peter2022end}. Speech recognition studies can use either time-domain or frequency-domain inputs. However, deep learning based on frequency-domain inputs has been proven to be a better approach if using a complex spectrogram to process audio signals \cite{nossier2020comparative}. Mel-Frequency Cepstral Coefficients (MFCCs) used to be a viral type of speech signal feature. Recently, filter banks are found to be more attractive \cite{mukherjee2019spoken}. Various audio-related deep learning models have widely used the Mel-spectrogram that applies a frequency-domain filter bank to audio signals (e.g., \cite{ gong2021ast}).

Speakers have unique voices and speaking habits. Therefore, speech recognition models are classified into user-dependent and user-independent models \cite{gaikwad2010review}. A dependent model is created for one particular speaker, whereas an independent model is for various speakers. Although a dependent model is easier to develop, it is a nontransferable point solution. Maintaining many point solutions is challenging in some real-world applications that have multiple model users or users that can change quickly. Therefore, user-independent models are the mainstream.

\subsection{Speaker Recognition}
\label{subsec:Lit Speaker Recognition}

Instead of recognizing spoken words, speaker recognition focuses on distinguishing speakers by their voices. As one of the biometrics to distinguish people, voice has a high ease of use and implementation, high user acceptance, and low cost compared to other higher accurate biometrics such as fingerprint and retina \cite{hanifa2021review}. Speaker recognition has two research streams, which are speaker verification and identification. Speaker verification aims to test if a given voice is from an authorized speaker. D-vector is a popular speaker representation that is derived from a deep neural network for speaker verification \cite{heigold2016end}. The d-vector of a test utterance is compared to a speaker model that is generally given by averaging over the d-vectors of the enrollment utterances. The cosine similarity is a popular similarity measurement for the d-vector comparison. A general speaker verification process is to compare the given voice to the enrollment voices and make an accept or reject decision according to a threshold. Speaker identification is to identify the speaker among a set of known speakers from the given test voice.

Before the era of deep neural networks, Gaussian mixture models \cite{reynolds1995robust} and i-vector \cite{dehak2010front} were popularly used in speaker recognition. The main phases of speaker recognition generally include pre-processing, feature extraction, modeling, and classification. In the recent decade, motivated by the success of deep learning, many deep neural networks for speaker recognition investigated not only classification (e.g. \cite{kenny2014deep}) but feature extraction, such as the d-vector method \cite{variani2014deep}. Feature extraction is the key to speaker recognition as it is supposed to extract sufficient features of the input signal for better modeling \cite{chaudhary2017feature}. \cite{bai2021speaker} summarized methods of deep speaker feature extraction. Converting raw data into acoustic features and turning them into deep embeddings using CNN-based deep feature extractors is becoming a major approach \cite{zhang2018text,yadav2018learning,wang2020multi,garcia2020jhu}. Popular acoustic features include spectrogram, Mel-spectrogram, and MFCC. The CNN-based feature extractors are diverse, for example, TDNN, ResNet, VGG, and Inception-ResNet. The study by \cite{chung2018voxceleb2} showed that the best performance on their test data is RestNet50, with an error rate of 4.42\%.

Past research observed a significant drop in speaker recognition performance when the number of speakers increased. For example, the accuracy is 93.7\% when the pool has 16 speakers, but it drops to 81.7\% when the pool size becomes 20 \cite{parveen2000speaker}. The accuracy drops from 96\% to 65.3\% when the speaker increases from 5 to 30 \cite{chauhan2017speaker}. \cite{chauhan2019speaker} proposed an artificial neural network combining three different feature types, which can keep the accuracy around 93\% when the speaker number increases from 20 to 30. 
Recently, deep neural networks trained on large-scale datasets have achieved high accuracy. For example, each of 630 speakers provided six phonetically rich sentences, and the data were used to train a model that achieved 97.0\% accuracy \cite{lukic2016speaker}. \cite{ye2021deep} trained the model using 127,551 utterances collected from 400 speakers, and the accuracy is 98.96\%.

\subsection{Multi-tasking Models Attained by Joint Training}

Keyword classification and speaker recognition have been considered two related tasks, not only because speech commands contain phonetic features and speaker-identity information but because integrating these tasks in a unified model is beneficial. The association of the two tasks arises from various real-world applications. The study by \cite{sigtia2020multi} was motivated by the need to detect voice-triggering phrases and verify if the speaker is a registered user. \cite{el2019joint} aimed to recognize who says what and when in a conversation setting. Personalized devices, such as hearing assistive devices, require the ability to detect external speakers and prevent them from triggering the device. \cite{lopez2020improved} developed a multi-tasking keyword spotting model with the ability to detect non-users.

Joint training of keyword and speaker recognition as a unified multi-tasking model usually has the following one or both benefits over two independent models. First, the two tasks can share the data processing and feature learning pipeline to some extent \cite{sigtia2020multi,lopez2020improved,tang2016collaborative,jung2020multi, hussain2022multi}. Second, each can benefit from the improved performance of the other \cite{tang2016collaborative, jung2020multi}. In the study of \cite{sigtia2020multi}, two stacks of four LSTM layers perform the voice trigger detection and the speaker verification tasks, respectively. The two tasks share the first two layers without sacrificing the accuracy compared to that achieved by two independent models. The two tasks in \cite{hussain2022multi} share the same wav2vec v2 backbone, and those in \cite{lopez2020improved} share a residual deep learning network for feature extraction. Through collaborative training, \cite{tang2016collaborative} developed a multi-tasking model for speech and speaker recognition. The two tasks share a common front-end but have their respective recurrent neural networks. The two networks are connected at the task level to inform each other of the desired and undesired information. The keyword spotting and speaker verification tasks in \cite{jung2020multi} share an enhancement network for noise removal. The two tasks have their respective feature extractors, but the acoustic feature extractor provides the phonetic conditional vector to augment the speaker feature extractor's ability. A pooling network further integrates outputs from the two feature extractors to generate the keyword and speaker embeddings.

Deep neural networks suffer from the catastrophic forgetting problem in class-incremental learning. This problem also challenges speech recognition, for example, the incremental classes of new accents, new words, or new acoustic environments \cite{fu2021incremental}. \cite{li2020automatic} and \cite{anand2019few} proposed few-shot learning-based speaker identification networks to handle new speakers. The effectiveness of few-shot learning for adding new speakers in a speaker-keyword classification multi-tasking model is awaiting verification.

\section{The Model}
\label{sec:Model}

\begin{figure}[htbp]
\centering
\includegraphics[width=0.8\columnwidth]{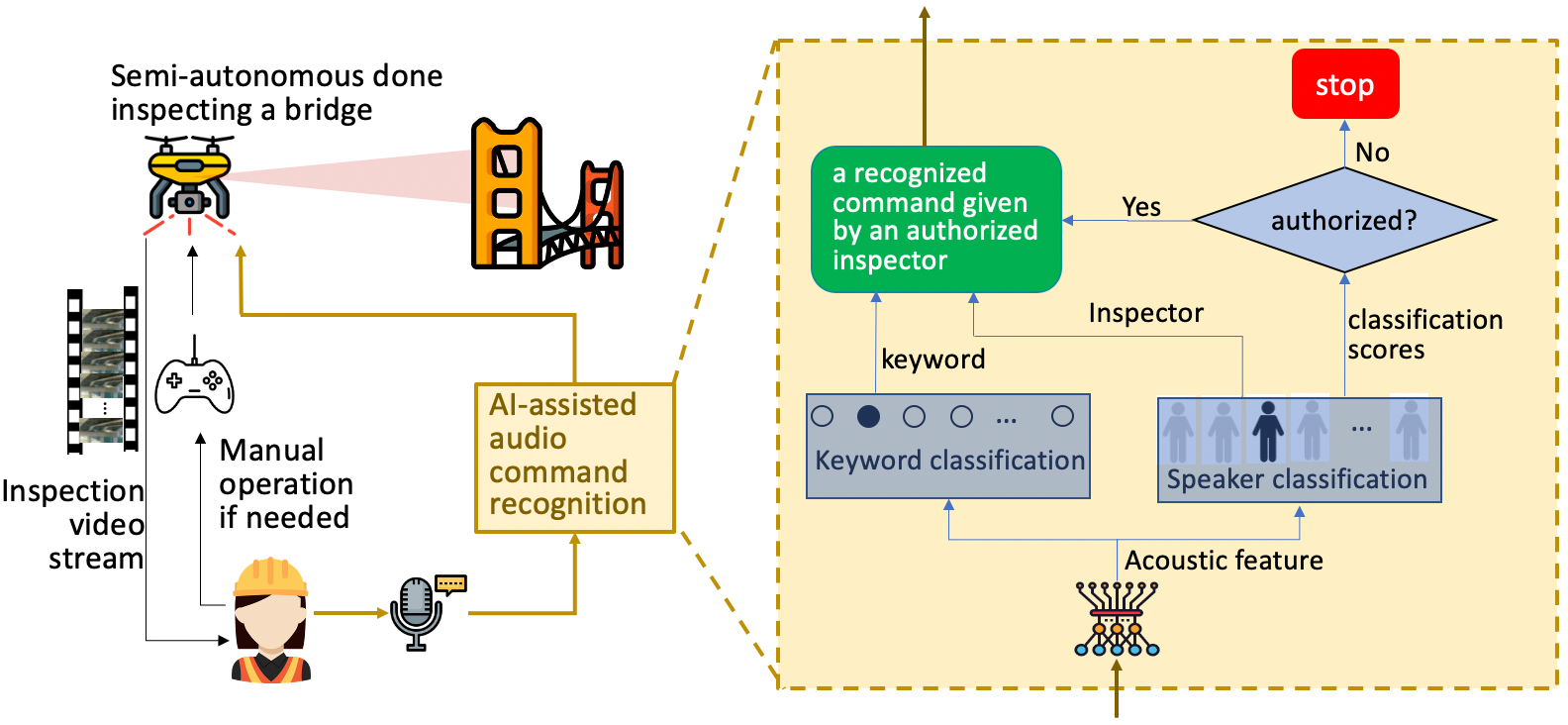}
	\caption{Collaborative human-robot inspection of bridges enabled by an AI model for inspector and audio command recognition.}
	\label{fig:general_diagram}
\end{figure}

Figure \ref{fig:general_diagram} describes the desired AI model to append to the human-robot system for bridge inspection. With that model, an authorized inspector can guide or assist a semi-autonomous drone in operation using audio commands. Nevertheless, the inspector can always take over the control by operating the drone manually using a hand controller. 

\subsection{The Rationale of the Proposed Model Architecture}
A simple method to build the desired capability of audio command recognition in Figure \ref{fig:general_diagram} is to collect sample keywords spoken by an inspector and build a model to classify the keywords and verify if the speaker is the authorized inspector. However, the model developed for one inspector cannot be transferred to others for multiple reasons. In the sample collected from an inspector, phonetic-specific features of keywords may be incomplete and mixed with other features that correlate to keyword classes but are unique to the inspector. For example, an inspector always speaks one keyword quickly and the other very slowly. Speed can be a dominating feature in distinguishing the two keywords that the inspector speaks. However, no other inspectors follow the same pattern of speaking. The classification model learned from such features is highly effective for that particular inspector but not others. Therefore, every inspector has to have a dedicated model developed using the inspector's sample data, termed a point solution in this paper. When the inspector is no longer an active worker, the inspector's model and data become obsolete. A method to address the mentioned limitations would be training a unified model on a dataset collected from multiple inspectors. In such a dataset, the sample of the same keyword has more considerable inter-speaker variability than intra-speaker variability. A portion of the inter-speaker variability positively contributes to speaker classification, and another portion benefits keyword classification by forming richer phonetic-specific features. If used appropriately, the pooled data allow for extracting richer keyword-specific features, learning speaker-identity features, and separating these reliably. From this perspective, training a unified classification model for a group of inspectors using the pooled data best utilizes data than training one dedicated model for each inspector.

In order to learn and attain the ability to classify both speakers and the keywords they spoke from the pooled training data, the model architecture must be appropriate. While the two classification tasks can have their respective feature extractors, sharing the feature extractor would be more efficient. Inspectors' acoustic signals contain rich information (e.g., phonetic features, pitch, intonation, rhythm, and accent) that can differentiate the keywords they speak and the speakers themselves. Therefore, a powerful feature extractor, like one based on deep learning, can automatically learn comprehensive features useful for both speaker and keyword classifications. Then, the two downstream tasks learn to use the extracted features selectively to achieve their respective missions.
Furthermore, jointly training the two classifiers would be necessary when the two tasks share a feature extractor. The back-propagation process will refine the feature extractor to be able to render the features that best support both tasks, not just a particular one. In addition, two mapping functions may be able to disentangle phonetic-specific and subject-specific features intertwined in the extracted features for the two classification tasks. From the model development perspective, including those two mapping functions may not always be necessary because the two classifiers can be trained directly on the extracted features. However, disentangling the extracted features for the two classifiers would make it easier to maintain the keyword classifier, update the speaker classifier to have incremental classes, and explicitly monitor the split effectiveness for any individual inspector in testing.

A single-user model can simply measure the similarity between an input utterance with enrollment samples of the authorized user to verify the speaker. However, speaker verification for the proposed multi-user model cannot follow the same method, especially when the group size of authorized inspectors is relatively large. Speaker classification scores should be a good reference for speaker verification if the classification result is reliable. Inter-subject variability is always present in people's voices and speaking habits, even for twins. Therefore, the speaker classification model tends to be less confident about classifying an unauthorized speaker into an authorized one. Nevertheless, the classification scores for authorized but hard-to-classify inspectors may have a similar pattern, and a method is needed to distinguish them from unauthorized speakers.

The above-discussed rationale for the modeling approach motivates the development of a unified multi-tasking deep learning model with a Share-Split-Collaborate (S$^2$C) learning architecture trained on pooled data. The remainder of this section further presents the speaker-keyword classification model and the speaker verification method.

\subsection{The Multi-tasking Classification Model}
\label{subsec:The Multi-tasking Model Architecture}

\begin{figure*}[htb]
\centering
\includegraphics[width=\columnwidth]{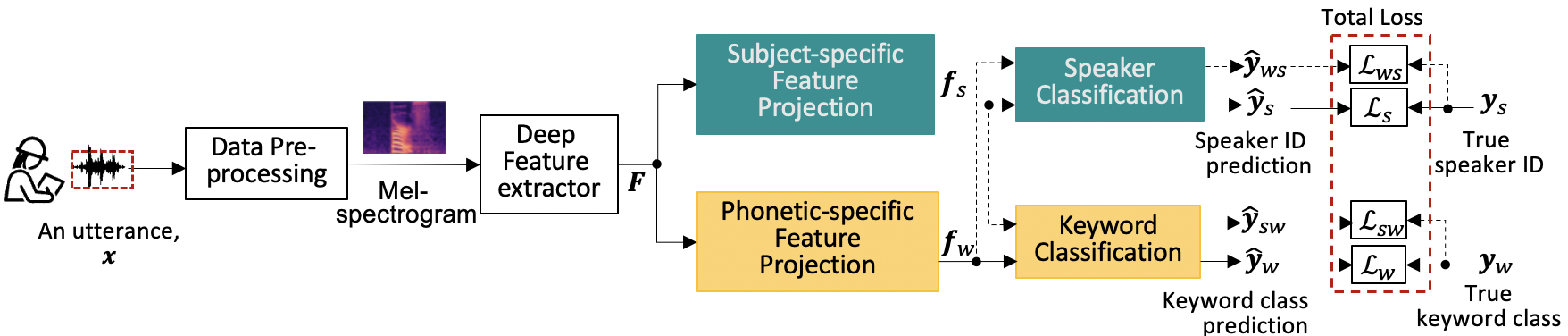}
\caption{The proposed Share-Split-Collaborate (S$^2$C) multi-tasking framework for the speaker-keyword classification}
\label{Fig:S3 Architecture}
\end{figure*}

Suppose a group of $M$ inspectors, indexed by their identification (ID) number $i$, will use one unified model to communicate with a drone for inspection using a set of $N$ keywords indexed by $j$. As Figure \ref{Fig:S3 Architecture} shows, an input to the model is an utterance that lasts for a fixed time period, $\pmb{x}$. The input utterance $\pmb{x}$ may come from one of the $M$ inspectors, which is indicated by a one-hot encoded vector, $\pmb{y}_s(\in\mathbb{R}^M)$. That is, at most one element of $\pmb{y}_s$ is one and all else is zero. If it exists, the element one indicates who the inspector is. $\pmb{x}$ may pertain to one of the $N$ keywords, represented by a one-hot encoded vector, $\pmb{y}_w\in\mathbb{R}^N$. Given an input $\pmb{x}$, the model predicts the speaker ID, $\widehat{\pmb{y}}_s$, and the keyword class, $\widehat{\pmb{y}}_w$, in parallel.

Figure \ref{Fig:S3 Architecture} shows the proposed S$^2$C deep learning architecture for the multi-tasking classification model. The framework contains a data pre-processing module, a deep feature extractor, two feature projection networks, two classification networks, and four loss functions. Details of these components and their relationship are introduced below.

\subsubsection{Shared Data Pre-processing and Deep Feature Extractor}
A data pre-processing module converts each input utterance $\pmb{x}$ into a Mel-spectrogram in size $3\times 224\times 224$, a frequency-domain representation of the raw input data. The deep feature extractor is ResNet50 \cite{he2016deep},  pre-trained on the ImageNet and transferred in for the speaker-keyword classification tasks. ResNet50 extracts a feature map, $\pmb{F}(\in  \mathbb{R}^{2048\times 7\times 7})$, from the Mel-spectrogram.

\subsubsection{Feature Disentanglement}
A feature projection network in equation (\ref{eq:projectionsubject}) attempts to split the subject-specific feature vector, $\pmb{f}_s$, from the extracted feature map $\pmb{F}$:
\begin{equation}
    \pmb{f}_s=L(\pmb{W}_{s,p}\pmb{F}+\pmb{b}_{s,p}),
    \label{eq:projectionsubject}
\end{equation}
where the transformation matrix $\pmb{W}_{s,p} (\in\mathbb{R}^{2048\times 2048})$ and the bias vector $\pmb{b}_{s,p}\in(\mathbb{R}^{2048})$ are the learnable parameters of the subject-specific feature projection network, and $L$ reshapes the output feature map into the feature vector $\pmb{f}_s(\in\mathbb{R}^{100352})$.

Similarly, the other projection network in equation (2) extracts the phonetic-specific feature vector, $\pmb{f}_w(\in\mathbb{R}^{100352})$, from $\pmb{F}$:
\begin{equation}
    \pmb{f}_w=L(\pmb{W}_{w,p}\pmb{F}+\pmb{b}_{w,p}),
    \label{eq:projectionkeyword}
\end{equation}
where $\pmb{W}_{w,p} (\in\mathbb{R}^{2048\times 2048})$ and $\pmb{b}_{w,p}\in(\mathbb{R}^{2048})$ are learnable parameters of the phonetic-specific feature projection network.

\subsubsection{Speaker and Keyword Classifiers}
Using the obtained subject-specific feature vector $\pmb{f}_s$ as the input, a network $\mathcal{N}_s$ in equation (\ref{eq:speakerclassifier}), named speaker classifier, performs the classification task and yields the probabilistic prediction of speaker ID, $\widehat{\pmb{y}}_s$. Specifically, the input $\pmb{f}_s$ passes through two fully-connected layers and an output layer in sequence to become the output $\widehat{\pmb{y}}_s$:
\begin{equation}
    \widehat{\pmb{y}}_s=\mathcal{N}_s(\pmb{f}_s;\phi_{s,1},\phi_{s,2},\gamma_{s}).
    \label{eq:speakerclassifier}
\end{equation}

$\phi_{s,1}$ in equation (\ref{eq:speakerclassifier}) is the first fully-connected layer performing the following operation:
\begin{equation}
 \pmb{h}_{s,1}=\text{ReLU}(\pmb{W}_{s,1}\pmb{f}_s+\pmb{b}_{s,1}),
\end{equation}
where the weight matrix $\pmb{W}_{s,1}(\in  \mathbb{R}^{128\times100352})$  and the bias vector $\pmb{b}_{s,1}(\in\mathbb{R}^{128})$ are the learnable parameters, the activation function is a ReLU function, and the output is $\pmb{h}_{s,1}$.

$\phi_{s,2}$ in equation (\ref{eq:speakerclassifier}) is the second fully-connected layer that performs the following operation:
\begin{equation}
 \pmb{h}_{s,2}=\text{S}(\pmb{W}_{s,2}\pmb{h}_{s,1}+\pmb{b}_{s,2}),
\end{equation}
where $\pmb{W}_{s,2}(\in  \mathbb{R}^{256\times128})$ and $\pmb{b}_{s,2}(\in\mathbb{R}^{256})$ are the learnable parameters, the activation function S is a sigmoid function, and $\pmb{h}_{s,2}$ is the output.

$\gamma_s$ in equation (\ref{eq:speakerclassifier}) is the output layer that turns the output from the second fully-connected layer $\pmb{h}_{s,2}$ into the probabilistic prediction of speaker ID $\widehat{\pmb{y}}_s$ with the following operation:
\begin{equation}
    \widehat{\pmb{y}}_s=\sigma(\pmb{W}_{s,o}\pmb{h}_{s,2}+\pmb{b}_{s,o}),
\end{equation}
where the weight matrix $\pmb{W}_{s,o} (\in\mathbb{R}^{M\times 256})$ and the bias vector $\pmb{b}_{s,o}(\in\mathbb{R}^{M})$ are learnable parameters, and $\sigma$ designates the soft-max function that normalizes the prediction scores as the probabilistic prediction.

The keyword classifier, $\mathcal{N}_w$, is similarly defined in equation (\ref{eq:keywordclassifier}), which turns the phonetic-specific feature vector $\pmb{f}_w$ into the probabilistic prediction of keyword class, $\widehat{\pmb{y}}_w$:
\begin{equation}
    \widehat{\pmb{y}}_w=\mathcal{N}_w(\pmb{f}_w;\phi_{w,1},\phi_{w,2},\gamma_{w}).
    \label{eq:keywordclassifier}
\end{equation}
where the first fully-connected layer $\phi_{w,1}$ is:
\begin{equation}
 \pmb{h}_{w,1}=\text{S}(\pmb{W}_{w,1}\pmb{f}_w+\pmb{b}_{w,1}),
\end{equation}
the second fully-connected layer $\phi_{w,2}$ is:
\begin{equation}
 \pmb{h}_{w,2}=\text{S}(\pmb{W}_{w,2}\pmb{h}_{w,1}+\pmb{b}_{w,2}),
\end{equation}
and the output layer $\gamma_w$ is:
\begin{equation}
    \widehat{\pmb{y}}_w=\sigma(\pmb{W}_{w,o}\pmb{h}_{w,2}+\pmb{b}_{w,o}).
\end{equation}
Here, $\pmb{W}_{w,1}(\in  \mathbb{R}^{512\times100352})$, $\pmb{b}_{w,1}(\in\mathbb{R}^{512})$, $\pmb{W}_{w,2}(\in  \mathbb{R}^{512\times512})$, $\pmb{b}_{w,2}(\in\mathbb{R}^{512})$, $\pmb{W}_{w,o}(\in\mathbb{R}^{N\times512})$, and $\pmb{b}_{w,0}(\in\mathbb{R}^{N})$ are the learnable parameters of the keyword classifier.

If the projection network in equation (\ref{eq:projectionsubject}) is effective, the subject-specific feature vector $\pmb{f}_s$ should be keyword-agnostic. That is, by entering $\pmb{f}_s$ to the keyword classifier, the obtained output $\widehat{\pmb{y}}_{sw}$,
\begin{equation}
    \widehat{\pmb{y}}_{sw}=\mathcal{N}_w(\pmb{f}_s;\phi_{w,1},\phi_{w,2},\gamma_{w}),
    \label{eq:ysw}
\end{equation}
does not tell what keyword it is. That is, the ideal prediction result for $\widehat{\pmb{y}}_{sw}$ is a uniform distribution $[\frac{1}{N},\dots,\frac{1}{N}]$.

To check if the phonetic-specific feature vector $\pmb{f}_w$ is speaker-agnostic, it can be entered into the speaker classifier to compare the output $\widehat{\pmb{y}}_{ws}$:
\begin{equation}
    \widehat{\pmb{y}}_{ws}=\mathcal{N}_s(\pmb{f}_w;\phi_{s,1},\phi_{s,2},\gamma_{s}),
    \label{eq:ysw}
\end{equation}
with its ideal result $[\frac{1}{M},\dots,\frac{1}{M}]$. The data flows for predicting $\widehat{\pmb{y}}_{sw}$ and $\widehat{\pmb{y}}_{ws}$ in Figure \ref{Fig:S3 Architecture} are dashed arrows, meaning that they are auxiliary, facilitating in modeling training and testing.

\subsubsection{The Loss Function for Collaborative Training}
The training dataset,  $\Omega_{\text{T}}=\{\pmb{x}(k), \pmb{y}_{s}(k), \pmb{y}_{w}(k)|k=1,\dots,K\}$, contains $K$ observations, where $\pmb{x}(k)$ is the input utterance indexed as $k$, $\pmb{y}_s(k)$ is the one-hot encoding of the truth speaker ID, and $\pmb{y}_w(k)$ is the one-hot encoding of the truth keyword class for $\pmb{x}(k)$. The proposed model predicts the speaker ID $\widehat{\pmb{y}}_s(k)$ and the keyword class $\widehat{\pmb{y}}_w(k)$. The goal of model training is to fit the deep feature extractor, the two projection networks, and the two classifiers, which is achieved by minimizing the loss function, $\mathcal{L}$, in equation (\ref{eq:totalloss}):
\begin{equation}
\mathcal{L}=\mathcal{L}_s+\mathcal{L}_w+\mathcal{L}_{sw}+\mathcal{L}_{ws},
\label{eq:totalloss}
\end{equation}
which consists of four components.

$\mathcal{L}_{s}$ in equation (\ref{eq:totalloss}) is a cross-entropy loss measuring the inaccuracy in classifying speakers by the subject-specific feature vector,
\begin{equation}
\mathcal{L}_s=-\sum_{k=1}^K<\pmb{y}_{s}(k),\log \hat{\pmb{y}}_{s}(k)>,
\end{equation}
where $<,>$ stands for the inner product of two vectors.
$\mathcal{L}_w$ in equation (\ref{eq:totalloss}) is also a cross-entropy loss measuring the inaccuracy in classifying keywords by the phonetic-specific feature vector,
\begin{equation}
\mathcal{L}_w=-\sum_{k=1}^K<\pmb{y}_{w}(k), \log \hat{\pmb{y}}_{w}(k)>.
\end{equation}
$\mathcal{L}_{sw}$ in equation (\ref{eq:totalloss}) is the mean squared errors that regulates the subject-specific feature vector to be keyword-agnostic, and this measurement is based on the Euclidean distance between the prediction $\widehat{\pmb{y}}_{ws}$ and its ideal score:
\begin{equation}
\mathcal{L}_{sw}=\frac{1}{K}\sum_{k=1}^K\|\widehat{\pmb{y}}_{sw}(k)-1/N\|_2^2.
\end{equation}
Similarly, $\mathcal{L}_{ws}$ regulates the phonetic-specific feature vector to be subject-agnostic,
\begin{equation}
\mathcal{L}_{ws}=\frac{1}{K}\sum_{k=1}^K\|\widehat{\pmb{y}}_{ws}(k)-1/M\|_2^2.
\end{equation}

\subsection{Inspector Verification}
\label{subsec:Inspector Verification}

Given an input utterance $\pmb{x}$ ($\notin \Omega_{\text{T}}$), the classification model presented in Section \ref{subsec:The Multi-tasking Model Architecture} renders the speaker classification scores $\hat{\pmb{y}}_s$ that measure the probabilities of being any of the $M$ speakers. If the speaker classification is highly reliable, the result is useful information for speaker verification. Therefore, a verification module is further developed, which directly uses the speaker classification result to verify if the speaker is in the pool of authorized inspectors. That is, the speaker verification module uses $\widehat{\pmb{y}}_s$ to provide the verification result $\widehat{y}_v\in[\text{`Authorized'},\text{`Unauthorized'}]$.

In predicting the class of an unauthorized speaker, the speaker classifier is likely to render prediction scores $\widehat{\pmb{y}}_s$ less capable of distinguishing speakers. A measure, $\lambda_v$, defined as the ratio of the highest score $\widehat{y}_s^{(1)}$ to the second highest score $\widehat{y}_s^{(2)}$ of $\widehat{\pmb{y}}_s$,
\begin{equation}
\lambda_v=\hat{y}_s^{(1)}/\hat{y}_s^{(2)},
\label{eq:lambdav}
\end{equation}
quantifies the minimum relative strength of the top-ranked prediction score and infers the confidence of predicting the speaker as the one with the highest score $\widehat{y}_s^{(1)}$. $\lambda_v$ takes values within the range $[1, \infty)$. The larger the value, the stronger the belief in the top-scored prediction. A threshold must be defined appropriately to differentiate unauthorized speakers from authorized inspectors according to $\lambda_v$.

This paper proposes a simple method in equation (\ref{eq:lambda}) for defining a classification threshold, $\lambda$, based on the speaker classification results on the training dataset: 
\begin{equation}
\lambda=\frac{1}{K}\sum_{k=1}^K\frac{1}{\text{var}[\hat{\pmb{y}}_{s}(k)]}
\label{eq:lambda}
\end{equation}
where $\text{var}[\hat{\pmb{y}}_s(k)]$ designates the variance of $\hat{\pmb{y}}_s(k)$, the classification scores for the person spoken the input utterance $\pmb{x}(k)\in\Omega_{\text{T}}$. A small variance indicates difficulty in trusting the prediction to be the top scorer. Equation (\ref{eq:lambda}) indicates the threshold $\lambda$ is dependent on the pool of authorized inspectors. First, $\lambda$ is smaller if authorized inspectors are easier to classify. Second, a straightforward derivation further shows the threshold $\lambda$ takes a value within the range $[M+1+\frac{1}{M-1}, \infty)$, and the lower boundary of the threshold value $M+1+\frac{1}{M-1}$ is approaching $M+1$ as $M$ increases. That is, the lower boundary increases with the number of authorized inspectors in the pool. The explicit dependency of the classification threshold $\lambda$ on the pool of authorized inspectors makes the verification adaptive to various groups of inspectors in any inspection jobs.

By comparing the ratio value obtained in equation  (\ref{eq:lambdav}) to the threshold value defined in equation (\ref{eq:lambda}), the speaker verification result is determined:
\begin{equation}
    \widehat{y}_v=\left\{
    \begin{array}{ll}
      \text{`Authorized'},   & \text{if\;}\lambda_v\ge \lambda; \\
      \text{`Unauthorized'},   & \text{otherwise.}
    \end{array}
    \right.
\end{equation}

\section{Implementation Details}
\label{sec:ImplementationDetails}

The study collected data from a training program to implement the model proposed in section \ref{sec:Model}. This section discusses the details of the data collection, model training, and model adaptation. The dataset (in the format of Mel-spectrogram, spectrogram, and MFCC) and source codes are available for download from the project webpage \cite{SpeechData_Github}.

\subsection{The Data}

The study collected inspection command data spoken by eight subjects. They were receiving training to guide a semi-autonomous drone in performing a bridge inspection job that consists of four tasks using a virtual reality-based training system \cite{li2022virtual}. In this study, the drone can automatically perform four inspection tasks by flying along the pre-planned routes of the tasks with GPS-based navigation and fundamental obstacle avoidance functions. An inspector triggers the start and termination of tasks and guides the drone to have certain deviations from the pre-planned tasks. Ten keywords in three categories were collected and summarized in Table \ref{tab:keywords}. ``BIRDS'' is the drone's name used as the wake-up command for triggering the communication with the drone. The communication is on until a silence of over two seconds is detected. The inspector uses the command ``Task $i$'' ($i$= One, Two, Three, Four) to let the drone start a specific task. Therefore, five keywords fall in the category of assignment commands. Four additional single-word commands allow the inspector to modify the automatic inspection mode. The inspector can use the command ``Backward'' to ask the drone to move reversely along the pre-defined path. The drone will stay still if it receives the command ``Hover''. The drone will continue to perform the uncompleted task automatically if the inspector says the command ``Continue''. The command ``Stop'' will terminate the current task and let the inspector take control of the drone. This list of keywords is an example developed based on one inspection job. Any inspection job with unique specifications can have its commands created, which can flexibly include special words not found in daily communication.

\begin{table}[htbp]
\centering
\caption{The list of keywords in an inspection job}
\begin{tabular}{l|l}
\hline
Category& Keywords\\
\hline
Wakeup & ``BIRDS''\\
Assignment & ``Task'', ``One'', ``Two'', ``Three'', ``Four''\\
Adjustment & ``Backward'', ``Continue'', ``Hover'', ``Stop''\\
\hline
\end{tabular}
\label{tab:keywords}
\end{table}

\begin{figure}[htbp]
    \centering
    \includegraphics[width=0.6\columnwidth]{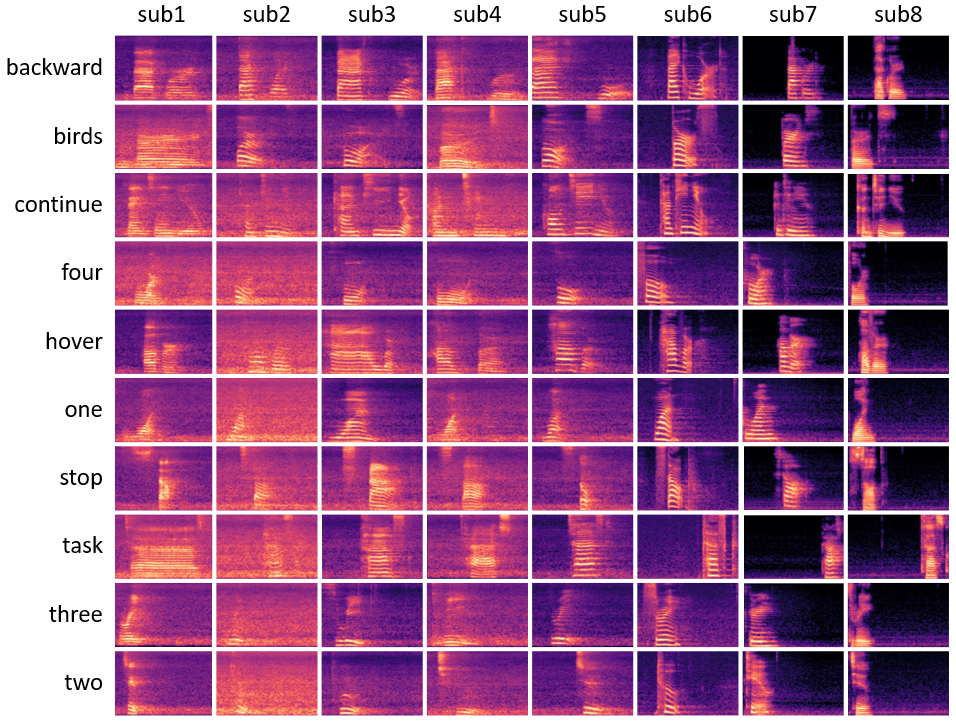}
    \caption{Mel-spectrogram samples of ten keywords spoken by eight subjects (sub1$\sim$sub8)}
    \label{fig:MEL}
\end{figure}

Half of the eight subjects in this study are female, and the other half are male. All subjects repeated each of the ten keywords about fifty times. Utterances of keywords were extracted from the recorded audio signals, each lasting 1.5 seconds. The utterances were all transformed into Mel-spectrogram images. Figure \ref{fig:MEL} illustrates one sample image of each keyword spoken by each of the eight subjects. The intra-row similarity and inter-row dissimilarly are observed, which evidences phonetic-specific features. The intra-row dissimilarity and the consistency across rows indicate that subject-specific features are present and phonetic-specific features from a single subject are probably incomplete.

\subsection{Model Training}

The model proposed in section \ref{subsec:The Multi-tasking Model Architecture} was trained, validated, and tested for a group of five subjects. The data is split into three subsets for three purposes: training (60\%, $\sim$30 utterances/keyword/subject), validation (20\%, $\sim$10 utterances/keyword/subject), and testing (20\%, $\sim$10 utterances/keyword/subject). Developing a satisfying model does not necessarily use up all the training data. Section \ref{subsec:Model Development Efficiency} will discuss the requirement for the training data size. The optimizer Adam was used for model training. The scheduler applies an exponential decay function to the optimization step, given a provided initial learning rate of 1e-4. The model training went with two stages. In the first stage, the feature extractor pre-trained on ImageNet was frozen, and the other four networks were trained from scratch for up to ten epochs. In the second stage, the feature extractor was unfrozen, and all five networks were refined for up to ten epochs. Each training stage may be terminated earlier if the validation accuracy is no longer improved for at least five epochs. The model that achieves the lowest loss on the validation dataset from the second stage is selected for use.

\subsection{Model Adaptation to Inspector Changes}
\label{subsec:Model Adaption to Worker Changes}

If any inspectors left the worker group (e.g., retirement and turnover), they become inactive users of the model. Regarding these changes, the speaker-keyword classification model does not require any change. A speaker recognized as an inactive inspector will not be able to communicate with the drone. However, updating the model is necessary if new inspectors join the group (e.g., new hires or contractors). The primary reason for updating the model is that the speaker classification network needs to handle new classes. Updating other model components (i.e., the feature extractor, the feature projection networks, and the keyword classification network) is unnecessary because they work for any individual. However, the additional training data collected from new inspectors may further improve the model's ability to extract features and classify keywords, mainly when the original training dataset contains only a small number of inspectors.

This study hypothesizes that incremental learning for adding new classes to the speaker classifier requires less data than training the initial model. Therefore, only five utterances of each keyword from each new inspector are added to the training data. Experiments in section \ref{subsec:Model Adaptability to New Inspectors}
validate that this small amount is sufficient. In the model calibration, the phonetic-specific feature projection network, the keyword classifier, and the feature extractor will use their current weights as the initial values. The underlying rationale is that those networks are at least near-optimal before the calibration, and the updated training dataset may help refine them to become optimal. However, the subject-specific feature projection network and the speaker classifier will use randomly assigned weights as the initial values. Re-training these two networks from scratch would avoid the issue of forgetting existing inspectors when learning to recognize new ones.

\section{Results and Discussion}
\label{sec:Results and Analysis}

This section presents nine experiments verifying the proposed model's functionality and merits. Results are analyzed and discussed below.

\subsection{Model Development Efficiency}
\label{subsec:Model Development Efficiency}
The first experiment studied the requirements on training data size and so the training time for achieving satisfying classification accuracy. The experiment consists of four speaker-keyword classification models trained and tested on the inspection keyword data collected from five subjects (sub1$\sim$sub5), which differ in the training data size (utterances/keyword). Table \ref{tab:trainingcost} compares the four models on the mean training time and mean classification accuracy based on 10 times of training and testing. The interval estimates are of 95\% confidence, and the small margins of error indicate the reported measurements are stable. The dataset for training the first model contains 10 utterances per keyword from each of the five subjects, indicated by a vector (10, 10, 10, 10, 10). The overall mean accuracy of keyword classification is 0.951$\pm$0.003, and the average accuracy at the subject-level ranges from 0.878 to 0.991. The overall mean accuracy of speaker classification is 0.969 $\pm$0.004, and the average accuracy of recognizing a specific subject ranges from 0.917 to 0.997. The model's ability to analyze subject \#2 is clearly lower than that for others. Subject \#2 seems quite different than others by looking at Figure \ref{fig:MEL}. To improve the accuracy in analyzing the audio commands for subject \#2, the second model is developed by doubling the training data. This time, the average accuracy in classifying the keywords for subject \#2 is effectively increased to 0.927, and the average accuracy of recognizing subject \#2 is increased to 0.987. If 0.95 is a desired accuracy level, the second model has not achieved a satisfying accuracy in classifying the keywords for subject \#2, which motivated the third model that added additional 10 utterances of each keyword from subject \#2 to the training data. This time, the keyword classification accuracy is at least 0.953 for every subject and the speaker recognition accuracy is at least 0.984. Compared to the fourth model that uses 30 utterances per keyword from all subjects, the third model is developed with less training data, shorter training time, but comparable performance. 

A series of two-sample t-tests assuming equal variances were performed to verify the observed changes in classification accuracy due to increases in the training data size. When the training data size increases from 10 utterances per keyword to 20, the changes in the mean accuracy in classifying keywords for the five individuals are evidenced by the p-values ranging from 0.000$\sim$0.017, and the changes in the mean accuracy in recognizing them are evidenced by the p-values ranging from 0.000$\sim$0.087. When extra 10 utterances per keyword from subject \#2 are further added to the training dataset, the change in the mean accuracy of keyword classification for this subject is statistically significant with the p-value equal to 0.000.

Therefore, in the remainder of the paper, the third model, trained using 30 utterances per keyword from subject \#2 and 20 utterances per keyword from the other four subjects, is used as the base model for further discussion. Training the base model is efficient because the 95\% confidence interval for the mean training time is 63$\pm$3 seconds. Transferring the pre-trained ResNet50 into this study as the deep feature extractor is an important reason for achieving time efficiency. The inference speed of the base model is reasonable, about 0.11 seconds per Mel-spectrogram image. Adding the required time to convert an utterance into a Mel-spectrogram, which takes about 0.03$\sim$0.04 seconds, the speed for analyzing acoustic signals is about 7 utterances per second.

\begin{table*}[htbp]
\centering
\caption{Impact of training data size on the mean training time and the mean accuracy of speaker-command classification}
\resizebox{\linewidth}{!}{%
\begin{tabular}{c|c|rrrrr|r|rrrrr|r}
\hline
Trn. Data Size&Estimate of Mean&\multicolumn{6}{c|}{Estimate of Mean Accuracy in Keyword Classification}&\multicolumn{6}{c}{Estimate of Mean Accuracy in Speaker Classification}\\
\cline{3-14}
\multicolumn{1}{c|}{(utterances/keyword)}&Trn. Time (sec)&sub1&sub2&sub3&sub4&sub5&\multicolumn{1}{c|}{overall}&sub1&sub2&sub3&sub4&sub5&\multicolumn{1}{c}{overall}\\
\hline
(10,10,10,10,10)&45±1&0.946&0.878&0.966&0.977&0.991&0.951±0.003&0.994&0.917&0.977&0.959&0.997&0.969±0.004\\	
(20,20,20,20,20)&57±3&0.970&0.927&0.988&0.988&0.999&0.974±0.003&0.991&0.987&0.992&0.987&1.000&0.991±0.002\\	
(20,30,20,20,20)&63±3&0.973&0.953&0.989&0.988&1.000&0.980±0.002&0.984&0.999&0.993&0.986&1.000&0.992±0.001\\	
(30,30,30,30,30)&74±4&0.985&0.958&0.989&0.988&1.000&0.984±0.002&0.982&0.994&0.997&0.990&1.000&0.993±0.001\\
\hline
\multicolumn{14}{l}{Note: The first column states the data size (utterances/keyword) collected from subjects \#1, \#2, \#3, \#4, and \#5, respectively.
}\\
\multicolumn{14}{l}{\hspace{2.3em} The value added after the sign of ``$\pm$'' stands for the margin of error, calculated based on ten repeats and at the level of significance 5\%.}
\end{tabular}
}
\label{tab:trainingcost}
\end{table*}

\subsection{Benefits of Pooled Training Data}

\begin{figure}[htbp]
\centering
	\includegraphics[width=\columnwidth]{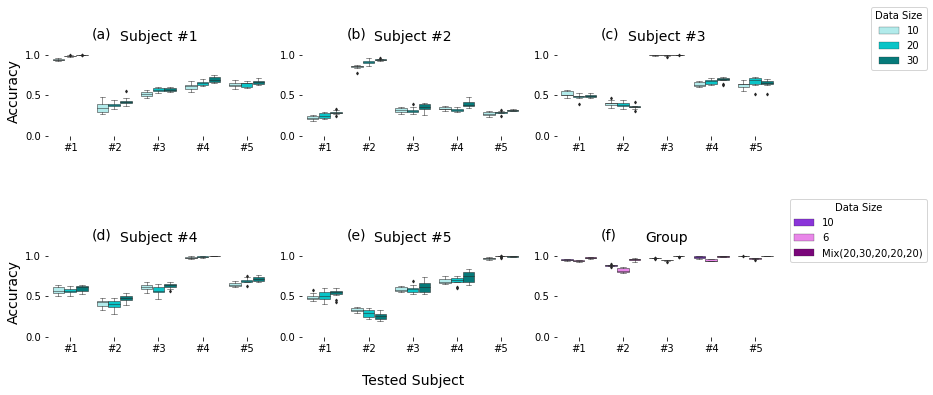}
	\caption{Comparison of models trained on one subject's data vs. on pooled data from a the group (training sample size are in utterances per keyword). }
	\label{fig:datapooling}
\end{figure}

The inter-subject variance of a keyword's feature vector is always present. Therefore, a model trained on a big volume of data collected from just one subject would not be effective for classifying the same set of keywords for other subjects. Pooling the data from a group of subjects has two advantages. On one hand, it allows for learning richer keyword representations. On the other hand, the inter-subject difference can be utilized to differentiate subjects. As a result, a unified model can be developed from the pooled data to substitute a set of point solutions that each is dedicated to one subject. The second experiment aims to verify the benefits of pooled training data, which consists of 18 models in 6 groups. The first column in Table \ref{tab:datapooling} shows the experiment design. The first 15 models (5 subjects $\times$ 3 levels of data size) are point solutions trained on the data collected from one subject and the last 3 models were trained on pooled data at 3 levels of data size. Each model was trained and tested 10 times. Table \ref{tab:datapooling} summarizes the average accuracy of keyword classification for each subject, and Figure \ref{fig:datapooling} compares the distribution of classification accuracy of those models using box plots.

Figure \ref{fig:datapooling}(a) presents the keyword classification accuracy of 15 models that each was trained on the data collected from subject \#1 and tested on all five subjects. The sizes of training data experimented with are 10, 20, and 30 utterances per keyword, respectively. Those box plots show that models trained on the data collected from subject \#1 only perform well on that subject, not others. When the training data size is 10 utterances per keyword, the average accuracy in classifying the keywords for subject \#1 is 0.937 but ranges from 0.341 to 0.630 for others. Increasing the sample size helps further improve the accuracy in classifying the keywords for subject \#1 (F-value for the one-way ANOVA test is 126.6), but not for others (F-values is less than 14.8). For example, when the training data size is tripled, the average accuracy in classifying the keywords for subject \#1 increases to 0.998, but the improvement for other subjects is 0.09 or less, far below a satisfying level. Observations from Figure \ref{fig:datapooling}(a-e) are consistent, indicating point solutions for individual subjects are non-transferable, regardless of the training sample size.

\begin{table}[htbp]
\centering
\caption{Effectiveness of pooled training data}
\begin{tabular}{l|rrrrr}
\hline
\multicolumn{1}{c|}{Trn. Data Size{$^\dagger$}}&\multicolumn{5}{c}{Avg. Accuracy of Keyword Classification$^{\ddagger}$}\\
\cline{2-6}
\multicolumn{1}{c|}{(utterances/keyword)}&sub1&sub2&sub3&sub4&sub5\\
\hline
(10,\;0,\;0,\;0,\;0)&0.937&0.341&0.515&0.604&0.630\\
(20,\;0,\;0,\;0,\;0)&0.983&0.378&0.565&0.649&0.633\\
(30,\;0,\;0,\;0,\;0)&0.998&0.428&0.570&0.694&0.663\\
\hline
(0,\,10,\;0,\;0,\;0)&0.219&0.848&0.319&0.335&0.272\\
(0,\,20,\;0,\;0,\;0)&0.246&0.911&0.311&0.319&0.286\\
(0,\,30,\;0,\;0,\;0)&0.286&0.938&0.346&0.392&0.316\\
\hline
(0,\;0,\,10,\;0,\;0)&0.516&0.396&0.993&0.635&0.622\\
(0,\;0,\,20,\;0,\;0)&0.481&0.380&0.993&0.668&0.665\\
(0,\;0,\,30,\;0,\;0)&0.492&0.358&0.992&0.691&0.647\\
\hline
(0,\;0,\;0,\,10,\;0)&0.567&0.414&0.609&0.975&0.643\\
(0,\;0,\;0,\,20,\;0)&0.562&0.397&0.574&0.984&0.691\\
(0,\;0,\;0,\,30,\;0)&0.598&0.471&0.627&0.990&0.711\\
\hline
(0,\;0,\;0,\;0,\,10)&0.488&0.328&0.583&0.688&0.963\\
(0,\;0,\;0,\;0,\,20)&0.502&0.289&0.588&0.686&0.989\\
(0,\;0,\;0,\;0,\,30)&0.532&0.252&0.616&0.733&0.986\\
\hline
(6,\;6,\;6,\;6,\;6)&0.935&0.821&0.940&0.941&0.966\\
(10,10,10,10,10)&0.946&0.878&0.966&0.977&0.991\\
(20,30,20,20,20)&0.973&0.953&0.989&0.988&1.000\\
\hline
\multicolumn{6}{l}{$\dagger$ The first column shows utterances/keyword from subjects \#1, 
}\\
\multicolumn{6}{l}{\#2, \#3, \#4, and \#5, respectively
}\\
\multicolumn{6}{l}{$\ddagger$ based on ten repeated training and testing
}\\
\end{tabular}
\label{tab:datapooling}
\end{table}

Figure \ref{fig:datapooling}(f) presents the keyword classification accuracy of the three unified models trained on the pooled data of the group and tested on each of the five subjects in the group. Three sample sizes were experimented with: (i) 6 utterances per keyword from each of the five subjects, (ii) 10 utterances per keyword from each subject, and (iii) 30 utterances per keyword from subject \#2 and 20 from all other subjects. The figure shows that the model trained on pooled data can classify all subjects’ keywords properly even when the training dataset contains only 6 utterances per keyword from each subject. Table \ref{tab:datapooling} shows that, when the training data size is 6 utterances per keyword per subject, the average accuracy in classifying the keywords for subject \#2 is 0.821 and at least 0.935 for other subjects. The performance of keyword classification could be largely improved by using more training data. The last model in Table \ref{tab:datapooling} is the base model, which achieves 0.953 or higher average accuracy in classifying the keywords at the subject level.

\subsection{Model Adaptability to New Inspectors}
\label{subsec:Model Adaptability to New Inspectors}

\begin{figure}[htbp]
\centering
	\includegraphics[width=0.9\columnwidth]{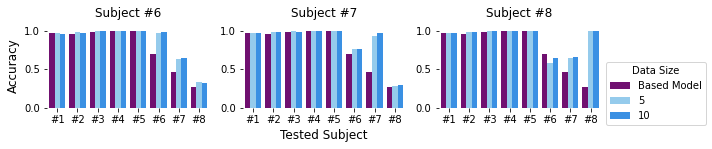}
	\caption{Keyword classification accuracy of the base model and the two adapted for a new subject using 5 vs. 10 utterances/keyword from the new subject. The base model is trained on the pooled data of subjects \#1$\sim$\#5, and subjects \#6$\sim$\#8 are new subjects.}
	\label{fig:owndata_adaptability}
\end{figure}

When a new inspector is joining a small group of inspectors, collecting a small amount of additional training data from the new inspector would be necessary to adapt the model to the new inspector. The third experiment of this study aims to evaluate the requirement on additional training data. The base model was respectively adapted for each of the three remaining subjects (i.e., sub6$\sim$sub8) using two sizes of training data: 5 and 10 utterances per keyword from the new subject. Results are summarized in Table \ref{tab:owndata_adaptability} and visualized in Figure \ref{fig:owndata_adaptability}. The base model's accuracy in classifying the keywords for subject \#6 is 0.700. When subject \#6 became a new inspector, the base model was calibrated by adding just 5 utterances per keyword collected from subject \#6 to the training dataset. The accuracy of the calibrated model in classifying the keywords for subject \#6 becomes 0.970, and the classification accuracy for the existing five subjects has no significant change. If subject \#7 is the new inspector, collecting 10 instances per keyword from the subject to calibrate the model effectively increases the accuracy from 0.464 to 0.968. Collecting 5 instances per keyword from subject \#8 can adapt the base model to this subject, manifested by the increase of the accuracy from 0.267 to 0.990. 
Figure \ref{fig:owndata_adaptability} shows that the additional 5 to 10 utterances per keyword from the new subject are also sufficient for the model to obtain the ability to recognize the new subject's spoken keywords with an accuracy near 1 without forgetting its ability to analyze the existing subjects.
The study verifies the efficiency and effectiveness of model adaptation for adding new inspectors.  Figure \ref{fig:owndata_adaptability} and Table \ref{tab:owndata_adaptability} also indicate that a model's ability to classify keywords for unauthorized subjects would increase if more subjects are included in the training data. For example, the base model's accuracy in classifying the keywords for subject \#8 is 0.267. After the base model is adapted to the bigger group that consists of subjects \#1$\sim$\#6, the accuracy of the updated model in classifying the keywords for subject \#8 is increased to 0.336. The observation indicates that phonetic-specific features learned on a big dataset with a large of group speakers should generalize well. Therefore, the model adaptation on a public dataset with more subjects is further examined in a later section.
\begin{table*}[htbp]
\centering
\caption{Model adaptation for adding new subjects: Keyword classification accuracy for existing and new subjects}
\begin{tabular}{c|c|rrrrr|rrr}
\hline
\multicolumn{2}{c|}{Additional Trn. Data}&\multicolumn{5}{c|}{Existing Subjects}&\multicolumn{3}{c}{New Subjects}\\
\hline
Source&(utterances/keyword)&sub1&sub2&sub3&sub4&sub5&sub6&sub7&sub8\\
\hline
\multirow{2}{*}{sub6}&5&0.973&0.981&0.991&1.000&1.000&0.970&0.632&0.336\\
&10&0.955&0.972&0.991&1.000&1.000&0.980&0.648&0.316\\
\hline
\multirow{2}{*}{sub7}&5&0.973&0.981&0.991&1.000&1.000&0.760&0.936&0.287\\
&10&0.973&0.981&0.982&1.000&1.000&0.760&0.968&0.297\\
\hline
\multirow{2}{*}{sub8}&5&0.964&0.981&0.991&1.000&1.000&0.580&0.648&0.990\\
&10&0.973&0.981&0.991&1.000&1.000&0.650&0.664&1.000\\
\hline
\multicolumn{2}{c|}{Base model}&0.973&0.963&0.982&0.990&1.000&0.700&0.464&0.267\\
\hline
\end{tabular}
\label{tab:owndata_adaptability}
\end{table*}

\subsection{Effectiveness of Inspector Verification}
\label{subsec:Effectiveness of Inspector Verification}
The study continued with the fourth experiment that used the base model to illustrate the effectiveness of the inspector verification method proposed in section \ref{subsec:Inspector Verification}. In this experiment, subjects \#1$\sim$\#5 are authorized inspectors, and subjects \#6$\sim$\#8 are unauthorized ones. The threshold $\lambda$ that equation (\ref{eq:lambda}) calculates for the base model is 7.048, close to its lower boundary of 6.25. If the ratio value $\lambda_v$ that equation (\ref{eq:lambdav}) calculates based on a speaker classification result is greater than $\lambda$, the inspector is predicted as an authorized inspector. The test data of the eight subjects are used to assess the performance of the inspector verification method. Table \ref{tab:speaker verification} summarizes the distribution of the ratio $\lambda_v$ for each subject. In total, 869 keyword utterances are tested, with 543 from the authorized inspectors and 326 from the unauthorized speakers. The statistic measurements of $\lambda_v$ in Table \ref{tab:speaker verification} clearly differentiate the two groups of subjects, verifying the rationale of using the ratio $\lambda_v$ as a measurement for inspector verification.
\begin{table}[htbp]
\centering
\caption{Statistics of the ratio value $\lambda_v$ for Inspector Verification}
\begin{tabular}{c|c|R{0.9cm}R{1cm}R{1cm}R{1.1cm}R{1.1cm}R{1cm}|R{1cm}R{1.1cm}}
\hline
&&\multicolumn{6}{c|}{$\lambda_{v}$ Distribution}&\multicolumn{2}{c}{1$\{\lambda_{v}>\lambda\}$}\\
\cline{3-10}
Subject ID&Test Sample Size&Min&Q1 & Q2 & Q3 &Max &Mean& Count&Percent \\
\hline
1&111 & 1.095 & 40.398 & 71.524 & 103.043 & 131.426 & 70.622 & 106 & 95.5\%\\
2&108 & 1.233 & 47.204 & 85.793 & 123.482 & 160.728 & 82.809 & 104 & 96.3\%\\
3&115 & 1.988 & 14.839 & 39.180 & 54.645 & 72.526 & 36.808 & 102 & 88.7\%\\
4&109 & 1.801 & 24.304 & 37.871 & 48.019 & 64.243 & 35.426 & 101 & 92.7\%\\
5&100 & 3.641 & 37.738 & 49.482 & 57.686 & 81.912 & 46.321 & 97 & 97.0\%\\
6&100 & 1.033 & 1.846 & 3.726 & 11.555 & 49.508 & 8.560 & 36 & 36.0\%\\
7&125 & 1.003 & 1.390 & 2.984 & 5.802 & 33.536 & 5.554 & 28 & 22.4\%\\
8&101 & 1.007 & 1.439 & 2.258 & 4.408 & 18.673 & 3.581 & 14 & 13.9\%\\
\hline
\end{tabular}
\label{tab:speaker verification}
\end{table}

Table \ref{tab:Inspector Verification} further presents the confusion matrix of inspector verification. 510 out of the 543 utterances from the authorized inspectors are predicted correctly, indicating the chance that the base model can correctly verify authorized inspectors is 93.9\%. 248 out of 326 utterances from the unauthorized inspectors are predicted correctly, which means the chance of successfully detecting an unauthorized speaker is 76.1\%. As a result, the precision of inspector verification is 86.7\% (=510/588), and the precision of unauthorized speaker detection is 88.3\% (=248/281). The result in Table \ref{tab:Inspector Verification} indicates the proposed inspector verification method is effective. A model user can use the threshold recommended by equation (\ref{eq:lambda}) as a starting point and adjust it according to the specific situation of implementation. For example, slightly increasing the threshold value $\lambda$ allows to increase the sensitivity of detecting unauthorized speakers, but it lowers the accuracy in verifying authorized inspectors. The lowered verification accuracy for authorized inspectors can be improved by adopting a temporal coherence analysis that verifies a speaker according to a sequence of acoustic inputs from the speaker rather than a single input.  

\begin{table}[htbp]
    \centering
        \caption{The confusion matrix of the base model's inspector verification result}
    \begin{tabular}{l|r|rr|r}
    \hline
        \multicolumn{2}{c}{}&\multicolumn{2}{c|}{Prediction}&\\
        \cline{3-4}
        \multicolumn{2}{c}{}&Authorized& Unauthorized &Total\\
        \hline
        \multirow{2}{*}{Ground Truth}&Authorized&510&33&543\\
        &\multicolumn{1}{c|}{Unauthorized}&78&248&326\\
        \hline
        \multicolumn{2}{r|}{Total}&588&281&869\\
        \hline
    \end{tabular}
    \label{tab:Inspector Verification}
\end{table}

\subsection{Impacts of Larger Group Sizes}

To demonstrate the applicability of the proposed speaker-keyword classification model to groups in larger sizes, an online audio dataset \cite{becker2018interpreting} is analyzed. This dataset includes 30,000 utterances of digits 0$\sim$9 collected from 60 different speakers who spoke every digit 50 times. The fifth experiment of this study, which is based on this spoken digit dataset, attempted to compare the keyword classification accuracy for authorized and unauthorized subjects and determine if having more authorized subjects in the training dataset would effectively reduce the difference between the two groups. Firstly, an initial model was developed for a group of 5 subjects. To keep consistent with the inspection command example in this paper, the dataset for training this classification model used the same data size. That is, the initial training dataset contains 20 utterances per keyword from each of the five subjects. The validation dataset and the test dataset respectively have 10 utterances per keyword from the subjects. 
The initial group of authorized subjects was expanded by adding one subject at once until reaching the size of 30 authorized subjects. When adapting the model to a new subject, 5 utterances per keyword from the subject were added to the training dataset to calibrate the model. In total, 26 models were developed, from a 5-subject group to a 30-subject group. Subjects whose data are included in the training dataset are called the authorized group and those whose data are not included are the unauthorized group. In this experiment, subjects 51$\sim$60 form a group of 10 unauthorized subjects. 

\begin{figure}[htb]
\centering
	\includegraphics[width=0.6\columnwidth]{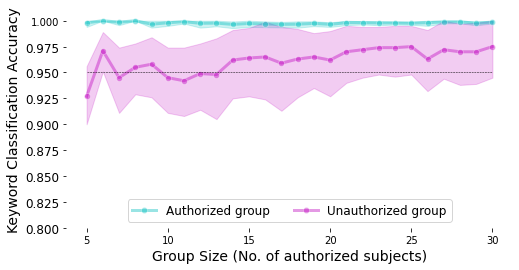}
	\caption{95\% confidence interval of the mean accuracy in classifying keywords at the individual level: authorized subjects vs. unauthorized subjects}
	\label{fig:digit_Data_performance}
\end{figure}

\begin{figure}[htbp]
\centering
	\includegraphics[width=0.6\columnwidth]{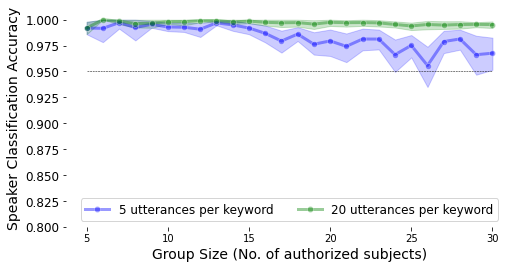}
	\caption{95\% confidence interval estimates of the mean accuracy in classifying keywords at the individual level: 5 vs. 20 utterances per keyword from each newly added subject}
	\label{fig:exsiting_performance}
\end{figure}

To verify that the keyword classification accuracy is not negatively affected by the increasing number of authorized subjects, the 95\% confidence intervals of each model's keyword classification accuracy for authorized subjects and unauthorized subjects are respectively shown in Figure \ref{fig:exsiting_performance}. The average accuracy in classifying the keywords spoken by the authorized subjects is near 1, and adding more and more subjects to the group does not change the average accuracy of keyword classification. The initial model's average accuracy in classifying the keywords spoken by unauthorized speakers is 0.927. It demonstrates a growing trend if the number of authorized subjects in the training dataset increases, reaching 0.975 when the size reaches 30 subjects. Subjects \#52, \#57, and \#60 are the three unauthorized subjects whose spoken keywords are classified by the initial model with 0.86$\sim$0.88 accuracy. When the number of authorized subjects is increased to 30, the mean accuracy in classifying the keywords of subjects \#57 and \#60 is 0.89, but it is 1 for subject \#52. The difference between the two groups' average accuracy is anticipated to diminish when the training dataset covers data from more and more subjects. However, the interval estimate of the mean classification accuracy for the unauthorized group is wider than that for the authorized group. Figure. \ref{fig:digit_Data_performance} confirms that pooling data from more subjects would reduce the gap of mean accuracy in keyword classification between the two groups, but not the difference in their variances. To achieve a reliable keyword classification result, it is recommended to include a small sample of training data from any new inspectors when the training dataset is not diverse enough.

The literature has shown that classifying speakers will become more challenging when the number of authorized inspectors keeps increasing. But this difficulty can be addressed by collecting more training data from the incrementally added subjects. The sixth experiment focused on verifying this hypothesis. Figure \ref{fig:exsiting_performance} shows the estimated mean value of subject-level speaker classification accuracy and the 95\% confidence interval under two scenarios: the training data contains 5 versus 20 utterances per keyword from each incrementally added subject. When only 5 utterances per keyword are collected from each sequentially added subject, the mean accuracy clearly demonstrates a decreasing trend when the number of authorized subjects increases, and the interval estimate becomes wider. If more training data are collected from newly added subjects, the decreasing trend of mean accuracy slows down and the interval estimate becomes narrower. Figure \ref{fig:exsiting_performance} implies that adapting the speaker-keyword classification model to a larger group of subjects may require more training data due to the increasing challenge of classifying speakers in a larger group.

\subsection{Assessment of the Modeling Approach}

This section illustrates additional experiments to compare the proposed modeling approach to alternatives in prior work, mainly from the aspects of data pre-processing, feature extraction, and speaker verification.

\subsubsection{Data Pre-processing Methods}
The literature review in section \ref{subsec:Lit Speaker Recognition} shows converting the raw data in the time domain into an acoustic feature in the frequency domain and then turning it into deep embedding using a convolutional neural network (CNN)-based deep feature extractor is gaining growing attention. The literature shows acoustic features in the frequency domain contain richer information than the raw data in the time domain. The seventh experiment compared three data pre-processing methods that respectively yield the following three types of acoustic features: Mel-Spectrogram, Spectrogram, and MFCC. Mel-Spectrogram is what the proposed model uses. Table \ref{tab:input_comparision} compares the use of these three pre-processed inputs by the base model (see section \ref{subsec:Model Development Efficiency}). The 95\% interval estimates of the mean training time and mean classification accuracy are based on ten times of model training and testing. Table \ref{tab:input_comparision} shows that different types of pre-processed inputs do not differentiate the mean training time, but Mel-Spectrogram surpasses Spectrogram and MFCC on the mean classification accuracy, manifested by an increase in the average keyword classification accuracy for 3 $\sim$ 3.5\% and an increase in the mean speaker classification accuracy for 4.1 $\sim$ 4.9\%.

\begin{table}[b]
\caption{Comparison of pre-processed acoustic features on 95\% confidence interval estimates of mean training time and mean classification accuracy}
\begin{tabular}{l|c|cc}
\hline
{ }                                     & \multicolumn{1}{l}{{ }}                                      & \multicolumn{2}{|c}{{ Mean Classification Accuracy}}  \\ \cline{3-4} 
\multirow{-2}{*}{{ Acoustic Feature}}    & \multicolumn{1}{l|}{\multirow{-2}{*}{{ Mean Training Time}}} & { Keywords} & { Speakers} \\ \hline
\multicolumn{1}{l|}{{ Mel-Spectrogram}} & \multicolumn{1}{c|}{{ 63±3}}                               & { 0.980±0.002}     & {0.992±0.001}     \\
\multicolumn{1}{l|}{{ Spectrogram}}     & \multicolumn{1}{c|}{{ 62±4}}                               & { 0.945±0.005}     & {0.944±0.004}     \\
\multicolumn{1}{l|}{{ MFCC}}            & \multicolumn{1}{c|}{{ 64±4}}                               & { 0.950±0.004}     & {0.951±0.004}     \\ \hline
\end{tabular}
\label{tab:input_comparision}
\end{table}

\subsubsection{Feature Extractors}
The eighth Experiment further compared three commonly used feature extractors in the literature: ResNet, VGG, and Inception-ResNet, on the base model. Table \ref{tab:feature_extractor_comparision} summarizes the 95\% confidence interval estimates of mean training time and mean classification accuracy based on 10 times of training and testing. ResNet50 is the feature extractor used in our proposed model, which achieves the highest average classification accuracy among the three feature extractors. Specifically, ResNet50’s average keyword classification accuracy is 6.8\% higher than VGG16 and 3.9\% higher than Inception-ResNet-v2. Its average speaker classification accuracy is 11.1\% higher than VGG and 9.7\% higher than Inception-ResNet-v2. The ResNet50’s average training time is comparable to VGG16 but about 30 seconds shorter than Inception-ResNet-v2. This experiment verifies the appropriateness of choosing ResNet50 as the feature extractor for the proposed classification model.
\begin{table}[htbp]
\caption{Comparison of feature extractors by the 95\% confidence interval estimates of mean training time and mean classification accuracy}
\begin{tabular}{l|c|cc}
\hline
{}                                     & \multicolumn{1}{l}{{ }}                                      & \multicolumn{2}{|c}{{ Mean Classification Accuracy}}  \\ \cline{3-4} 
\multirow{-2}{*}{{ Feature Extractor}}    & \multicolumn{1}{l|}{\multirow{-2}{*}{{Mean Training Time}}} & { Keywords} & { Speakers} \\ \hline
\multicolumn{1}{l|}{{ ResNet50}}                                                                            & \multicolumn{1}{c|}{{ 63±3}}                               & {0.980±0.002}     & {0.992±0.001}     \\
\multicolumn{1}{l|}{{ VGG16}}                                                                               & \multicolumn{1}{c|}{{ 59±2}}                               & {0.912±0.005}     & {0.881±0.004}     \\
\multicolumn{1}{l|}{{ Inception-ResNet-v2}}                                                                 & \multicolumn{1}{c|}{{ 93±6}}                               & {0.941±0.006}     & {0.895±0.007}     \\ \hline
\end{tabular}
 \label{tab:feature_extractor_comparision}
\end{table}

\subsubsection{Speaker Verification Methods}

Many studies found that the d-vector is an effective speaker representation extracted by a deep neural network for speaker verification. The d-vector of a test utterance is compared to a speaker model, generally the average over the d-vectors of the speaker’s enrollment utterances. The cosine similarity is a measurement for comparing the test d-vector to the speaker model. If the cosine similarity score exceeds a pre-specified threshold, the test utterance is verified as coming from the authorized speaker. The mechanism of the d-vector method indicates it is a suitable method to verify the speaker if the keyword classification model is developed for a single user. However, for a unified model with multiple authorized users, the d-vector method needs a speaker model for every authorized user and calculates the cosine similarity values between the test d-vector and each speaker model. For example, if the model is shared by 30 users, then 30 speaker models are required and 30 cosine measurements are calculated. The computational time increases linearly with the number of authorized speakers. Averaging all authorized users’ d-vectors to obtain the authorized speaker group model is reckless with poor performance. Section \ref{subsec:Inspector Verification} proposed a simple ratio method based on the speaker classification scores for the multi-user model, which was verified to be effective in section \ref{subsec:Effectiveness of Inspector Verification}.

\begin{figure}[htbp]
\centering
	\includegraphics[width=0.8\columnwidth]{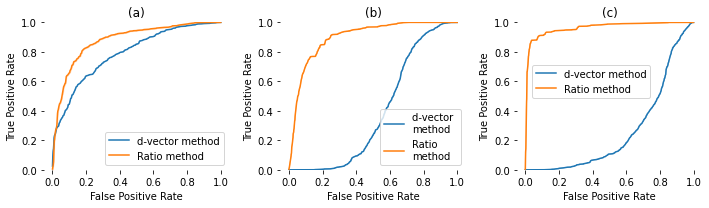}
	\caption{Receiver Operating Characteristic (ROC) curve for speaker verification on the spoken digit dataset: d-vector method vs. ratio method.}
	\label{fig:ROC}
\end{figure}

The ninth experiment compared the speaker verification performances of the d-vector method that is based on the authorized speaker group model to the proposed ratio method on three models: (a) the model trained for subjects \#1$\sim$\#5, and the training dataset contains 20 utterances per keyword per subject; (b) the model trained for subjects \#1$\sim$\#30, and the training dataset contains 20 utterances per keyword per subject for subjects \#1$\sim$\#5 and 5 utterances per keyword per subject for subjects \#6$\sim$\#30; and (c) the model is trained for subjects \#1$\sim$\#30, and the training dataset contains 20 utterances per keyword per subject. In this experiment, unauthorized speakers are defined as the positive class and authorized speakers as the negative class, assuming detecting unauthorized speakers is a high priority for a model user. Figure \ref{fig:ROC} compares the receiver operating characteristic (ROC) curves (i.e., true positive rate vs. false positive rate at various detection threshold values) of the d-vector method and the ratio method for each of the three models. AUC is the area under the ROC curve, which can take any value from 0 to 1. The larger the AUC value the better the performance in detecting unauthorized speakers. The AUC values of the d-vector method implemented by the three models are 0.7909, 0.3933, and 0.2608, while the AUC values of the ratio method are 0.8795, 0.8969, and 0.9623. That is, the d-vector method performs poorly when the group size is relatively large, but the ratio method is more reliable. The performance of the ratio method can be further improved by increasing the training data.

\section{Conclusions}
\label{sec:Conclusions}
This paper developed a unified multi-tasking deep learning model to classify the keywords of inspectors’ commands for guiding a semi-autonomous drone in inspection and, simultaneously, recognize inspectors who spoke the commands. The proposed S$^2$C architecture allows the two classification tasks to share the feature extractor and then split the subject-specific and keyword-specific features intertwined in the extracted features through feature projection and collaborative training. The model was trained on pooled data collected from the group of authorized inspectors who will use the model to stay in the loop of the semi-autonomous drone-assisted inspection.

This study collected an inspection keyword dataset from eight subjects to illustrate the proposed model. The dataset contains ten keywords that each of the subjects repeated every keyword about fifty times. A base model developed for a group of five authorized subjects achieved 95.3\% or higher mean accuracy in classifying the keywords for any authorized inspectors. Its mean accuracy in speaker classification is 99.2\%. The proposed model has effectively addressed the non-transferability of point solutions trained on the data of one subject and used only by that subject. Adapting the base model to include a new subject only requires collecting a small amount of training data from the new subject, like five utterances per keyword. The speaker prediction scores are effective for speaker verification. The base model's success rate in verifying authorized subjects is 93.9\%, and 76.1\% in detecting unauthorized ones. Increasing the training sample size can effectively improve speaker verification accuracy. The proposed model was further trained and tested on a public audio command dataset collected from sixty subjects. The proposed model was successfully applied to large-size groups, manifested by the consistently high and reliable keyword classification accuracy. Speaker classification will become more challenging when the group size is large, indicated by the decreased mean value and the enlarged variance of subject-level accuracy in speaker recognition. At the minimum, collecting sufficient training data from the subjects can address the challenge effectively.

Future work beyond this paper exists in three dimensions. First of all, implementing the classification model requires additional effort beyond this paper's study score. For example, an additional module is needed for locating command-related segments from stream data and extracting keyword utterances from the segments. Every working environment is unique, and every inspector is special. A research question is how to further improve the model's adaptability to noises in open-working environments and the varied spoken habits of inspectors. Effective data augmentation and model adaptation methods are possible solutions. Secondly, the current model can be further improved in multiple directions. The current study assumes a bijective relationship between the audio commands and the drone's actions. A more user-friendly approach to communication requires a surjective function that allows for mapping variants of each command from inspectors to the corresponding action of the drone. If new inspectors propose adding new keywords, class-incremental learning will happen to both classification tasks. Lastly, the study of this paper builds a methodological foundation for addressing the negative transfer issue, meaning a richer representation of speech is not learned from pooled data. Surface electromyography (sEMG) based silent speech recognition has a variety of critical applications, such as healthcare and defense, but it suffers from the negative transfer issue. Knowledge gained on multi-modal data (i.e., audio and sEMG) can be distilled to models that use sEMG as the single input for silent speech recognition. This paper provides an opportunity to explore those new research questions.

\section*{Acknowledgment}
Qin and Li are supported by National Science Foundation through the award ECCS-\#2026357.
Yao is supported by the SBU-BNL seed grant (1168726-9-63845) and National Science Foundation through the award ECCS-\#2129673.
\vspace{0.1in}

\bibliographystyle{plain}
\bibliography{references}
\end{document}